\documentclass[11pt,preprintnumbers,aps,amssymb,nofootinbib,amsmath,superscriptaddress,notitlepage,prd]
{revtex4-1}
\usepackage{epsfig,epsf}
\usepackage{bm} % puts greek and math symbols in boldface using \bm
\usepackage{color} % {\color{red} ... }
\definecolor{darkgreen}{RGB}{0,128,0}
\definecolor{darkred}{RGB}{139,0,0}
\usepackage{slashed} % Dirac slash
\usepackage{hyperref}
\newcommand{\beq}{\begin{equation}}
\newcommand{\beql}[1]{\begin{equation}\label{#1}}
\newcommand{\eeq}{\end{equation}}
\def\bal#1\gal{\begin{align}#1\end{align}}
\newcommand{\ball}[1]{\bal\label{#1}}
\newcommand{\eq}[1]{(\ref{#1})}

\renewcommand{\sec}[1]{Sec.~\ref{#1}}
\renewcommand{\b}[1]{{\bm #1}} 
\newcommand{\unit}[1]{\hat {{\bm #1}}} % unit vector

\newcommand{\im}{\,\mathrm{Im}\,}

\newcommand{\e}{\varepsilon}

\begin{document}

\title{Radiative instability of quantum electrodynamics in chiral matter}

\author{Kirill Tuchin}

\affiliation{Department of Physics and Astronomy, Iowa State University, Ames, IA 50011, USA}

\date{\today}


\begin{abstract}

Modification of the photon dispersion relation in chiral matter enables $1\to 2$ scattering.  As a result, the single fermion and photon states are unstable to photon radiation and pair production respectively.  In particular, a fast fermion moving through chiral matter can spontaneously radiate a photon, while a photon can spontaneously radiate a fast fermion and anti-fermion pair. The corresponding spectra are derived in the ultra-relativistic approximation. It is shown that the polarization of the produced and decayed photons is determined by the sign of the chiral conductivity. Impact of a flat thin domain wall on the spectra is computed.

\end{abstract}

\maketitle

%%%%%%%%%%%%%%%%%%%%%%%%%%%%%%%%%%%%%%%%
\section{Introduction}\label{sec:i}

One of the macroscopic manifestations of the chiral anomaly of QCD is the emergence of the topological $CP$-odd domains in hot nuclear matter \cite{Kharzeev:2007tn}. QED is coupled to these domains via its own chiral anomaly. This is represented by the triangular diagrams that involve two photon fields and the axial current generated by the topological fluctuations of the gluon field. The axial current rapidly increases with temperature which triggers a variety of non-trivial electromagnetic effects in quark-gluon plasma \cite{Kharzeev:2013ffa}.  

At a more fundamental level, the chiral anomaly makes photon effectively massive. Consequently, single photon and fermion states become unstable. Recall that  photon radiation by a charged fermion in vacuum $f(p)\to f(p')+\gamma(k)$ and the cross-channel process of pair production in vacuum $\gamma(k)\to f(p')+\bar f(p)$ are prohibited by momentum conservation.\footnote{$p$, $p'$ and $k$ are four-momenta with the components $p=(\e,\b p)$, $p'=(\e',\b p')$ and $k=(\omega, \b k)$.} Indeed  in the rest frame of one of the fermions $k^2= (p\pm p')^2= 2m(m\pm\e)$. The right-hand-side never vanishes since $\e>m$, whereas in the left-hand-side $k^2=0$ \cite{Berestetsky:1982aq}. In chiral matter, i.e.\ in a matter supporting the $CP$-odd domains,  the chiral anomaly modifies the photon dispersion relation  as \cite{Carroll:1989vb,Lehnert:2004hq,Kostelecky:2002ue,Tuchin:2014iua,Yamamoto:2015maz,Tuchin:2017vwb,Qiu:2016hzd}\footnote{In covariant form $k^2=-\lambda\sqrt{ (n\cdot k)^2-n^2k^2}$, where $n^\mu=\sigma_\chi\delta^\mu_0 $ in the matter rest frame.} 
\ball{i1}
k^2= -\lambda\sigma_\chi|\b k|\,,
\gal 
where $\lambda$ and $\b k$ are photon helicity and momentum and $\sigma_\chi$ is the chiral conductivity \cite{Fukushima:2008xe,Kharzeev:2009fn,Kharzeev:2009pj}. This opens the $1\to 2$ scattering channels, viz.\ the pair-production if $k^2>0$ and the photon radiation if $k^2<0$. Thus, single-particle states in chiral matter are unstable with respect to spontaneous radiation and decay. Moreover, in a matter with positive $\sigma_\chi$, only the right-polarized photons with $\lambda=+1$  can be radiated, while only the left-polarized photons with $\lambda=-1$ decay and vice-versa in a matter with negative $\sigma_\chi$.  

The main goal of this paper is to compute the photon radiation and pair production spectra due to the modified photon dispersion relation in chiral matter.  The fermions (and antifermions) are considered to be external particles propagating through the chiral matter. Their energy is assumed to be  much larger than the medium ionization energy.
The calculation method is borrowed from \cite{Schildknecht:2005sc} and 
relies on  several approximations: (i) photons and fermions are ultra-relativistic in the laboratory frame (the one associated with the matter), this means that $\e\gg m$, $\omega\gg \sigma_\chi$. Apart from making calculations significantly less bulky, this allows one to neglect the effect of the electromagnetic field instability in the infrared region as explained in \sec{sec:a}.  (ii) The matter is assumed to be spatially homogeneous and consisting of either one infinite   $CP$-odd domain  or of two semi-infinite domains separated by a flat domain wall.

The paper is structured as follows. In \sec{sec:a} and \sec{sec:b}  the wave functions of an ultra-relativistic  photon and fermion in chiral matter are derived.  In \sec{sec:d}  they are employed to compute the photon spectrum radiated  by a fermion in chiral matter with constant $\sigma_\chi$ as well as in matter with two chiral domains separated by a thin flat domain wall. The cross-channel process of pair production is analyzed in \sec{sec:e}. The results are summarized and discussed  in  \sec{sec:s}.

%\hrulefill

%%%%%%%%%%
\section{Photon wave function}\label{sec:a} 

%% see p. 8, 16, 22-24 of notes
The $CP$-odd domains in the chiral matter can be  described by a scalar field $\theta$ whose interaction with the electromagnetic field  $F^{\mu\nu}$ is governed by the Lagrangian \cite{Wilczek:1987mv,Carroll:1989vb,Sikivie:1984yz,Kalaydzhyan:2012ut}
\ball{a3}
\mathcal{L}= -\frac{1}{4}F_{\mu\nu}^2-\frac{c_A}{4}\theta\tilde F_{\mu\nu}F^{\mu\nu}+\bar \psi (i\gamma^\mu D_\mu-m)\psi\,,
\gal
where   $\tilde F_{\mu\nu}= \frac{1}{2}\epsilon_{\mu\nu\lambda\rho} F^{\lambda\rho}$ is  the dual field tensor and 
$c_A$ is the chiral anomaly coefficient \cite{Fujikawa:2004cx}.  A working assumption of this paper is that the field $\theta$ is spatially uniform (apart of possible thin domain walls) and a slowly varying function of time.  The ``free" field equations of electrodynamics in chiral electrically neutral and non-conducting matter read 
\label{b2}
\bal
&\b \nabla\cdot \b B= \b \nabla\cdot \b E=0\,, \label{b3}\\
& \b \nabla \times \b E= -\partial_t \b B\,,\label{b5}\\
& \b \nabla \times \b B= \partial_t \b E+\sigma_\chi \b B \,,\label{b6}
\gal
where $\sigma_\chi = c_A\dot \theta$ is the chiral conductivity \cite{Fukushima:2008xe,Kharzeev:2009fn,Kharzeev:2009pj}. (Often $\dot\theta$ is denoted by $\mu_5$ and is referred to as the axial chemical potential \cite{Huang:2015oca}).   
In the radiation gauge $A^0=0$ and $\b \nabla\cdot\b A= 0$ the vector potential obeys the equation
\ball{b8}
-\nabla^2\b A= -\partial_t^2\b A+\sigma_\chi (\b \nabla\times \b A)\,.
\gal
Its solution describing a photon moving along the $z$-direction with energy $\omega\gg k_\bot, \sigma_\chi$ is described by the wave function 
\ball{b10}
\b A^{(0)}= \frac{1}{\sqrt{2\omega V}}\,\b  e_\lambda  \,e^{i k_z z-i\omega t }\,,\quad k_z=\omega\,,
\gal
where the polarization vector satisfies  $\b  e_\lambda \cdot \unit z =0$. $V$ is the normalization volume. It is convenient to use the helicity basis $\b  e_\lambda =(\unit x+i\lambda\unit y)/\sqrt{2}$.  To determine the effect of the chiral anomaly on the photon wave function, look for a solution in the form
\ball{b12}
\b A= \frac{1}{\sqrt{2\omega V}}\,(\b  e_\lambda \varphi+\unit z \varphi') \,e^{i \omega z-i\omega t }\,,
\gal
where $\varphi$ and $\varphi'$ are functions of coordinates  slowly varying in the longitudinal ($z$) direction, viz.\ $|\partial_z\varphi/\varphi|\ll \omega $ and  $|\partial_z\varphi'/\varphi'|\ll \omega $. The two unknown functions $\varphi$ and $\varphi'$ are required in order to account for the change of the  photon polarization direction. The gauge condition yields a constraint
\ball{b13}
(\b e_\lambda \cdot \b \nabla_\bot) \varphi+\partial_z\varphi'+i\omega \varphi'\approx (\b e_\lambda \cdot \b \nabla_\bot) \varphi+i\omega \varphi'=0\,.
\gal
Substituting \eq{b12} into \eq{b8} one obtains
\ball{b14}
\b  e_\lambda \left(-2i\omega \partial_z\varphi-\nabla_\bot^2\varphi\right)+\unit z \left(-2i\omega \partial_z\varphi'-\nabla_\bot^2\varphi'\right)= \sigma_\chi\left(\omega \lambda \b e_\lambda \varphi - \b e_\lambda \times \b\nabla \varphi-\unit z\times \b\nabla_\bot \varphi'\right)\,.
\gal
Taking the scalar product of this equation with $\b  e_\lambda ^*$ and  using $\b e_\pm^*\cdot \b e_\mp=0$ and $\b e_\pm^*\cdot \b e_\pm=1$  produces
\ball{b16}
-2i\omega \partial_z\varphi-\nabla_\bot^2\varphi= \sigma_\chi (\omega\lambda \varphi-i\lambda \partial_z\varphi)+i\lambda\b e_\lambda ^*\cdot \b \nabla_\bot\varphi'\,,
\gal
where we used the identity $\b e_\lambda\times \unit z= i\lambda \b e_\lambda$.
In view of \eq{b13} one can drop the small term proportional to $\varphi'$. Neglecting also $\partial_z\varphi$ in parentheses furnishes an equation for $\varphi$:
\ball{b18}
-2i\omega \partial_z\varphi-\nabla_\bot^2\varphi= \sigma_\chi \omega\lambda \varphi\,.
\gal
Taking the scalar product of \eq{b14} with $\unit z$ yields
\ball{b20}
-2i\omega \partial_z\varphi'-\nabla_\bot^2\varphi'= \sigma_\chi i\lambda(\b e_\lambda \cdot \b \nabla)\varphi\,.
\gal
One can eliminate in \eq{b20} the term proportional to $\varphi$ using the gauge condition \eq{b13}. This furnishes an equation for $\varphi'$ which is precisely the same as equation \eq{b18} obeyed by $\varphi$. 

A solution to \eq{b18} can be written as
\ball{b22}
\varphi= e^{i\b k_\bot \cdot \b x_\bot}\exp\left\{ -i \frac{1}{2\omega} \int_0^z\left[k_\bot^2-\sigma_\chi(z')\omega \lambda\right]dz'\right\}\,.
\gal
It follows from \eq{b13} that 
\ball{b24}
\varphi' = -\frac{\b e_\lambda\cdot \b k_\bot}{\omega}\varphi\,.
\gal
Substituting \eq{b22} and \eq{b24} into \eq{b12} yields the photon wave function in the high energy approximation
\ball{b25}
\b A= \frac{1}{\sqrt{2\omega V}}\b  \epsilon_\lambda  \,e^{i \omega z+i\b k_\bot \cdot \b x_\bot-i\omega t }\exp\left\{ -i \frac{1}{2\omega} \int_0^z\left[k_\bot^2-\sigma_\chi(z')\omega \lambda\right]dz'\right\}\,,
\gal
where  the polarization vector 
\ball{b26}
\b\epsilon_\lambda = \b e_\lambda - \frac{\b e_\lambda\cdot \b k_\bot}{\omega} \unit z\,.
\gal
Clearly, $\b  \epsilon_\lambda \cdot \b k=0$ up to the terms of order $k_\bot^2/\omega^2$ and $\sigma_\chi/\omega$. 
 If the scattering process happens entirely within a single domain, then the chiral conductivity is constant. However, if a domain wall is located at, say, $z=0$, than the chiral conductivity is different at $z<0$ and $z>0$. This is why a possible $z$-dependence of $\sigma_\chi$ is indicated in \eq{b25}. Even though the boundary conditions on the  domain wall induce a reflected wave, it can be neglected in the ultra-relativistic approximation \cite{Baier:1998ej,Schildknecht:2005sc}.

It is seen in \eq{i1} that half of the infrared modes  $|\b k|<\lambda\sigma_\chi$ have $\im \omega>0$  implying  exponential growth  of the corresponding wave function with time. This infrared instability and its applications are discussed in many recent publications \cite{Joyce:1997uy,Boyarsky:2011uy,Tashiro:2012mf,Pavlovic:2016gac,Yamamoto:2016xtu,Xia:2016any,Manuel:2015zpa,Kharzeev:2013ffa,Khaidukov:2013sja,Kirilin:2013fqa,Avdoshkin:2014gpa,Akamatsu:2013pjd,Dvornikov:2014uza,Buividovich:2015jfa,Sigl:2015xva,Kirilin:2017tdh,Tuchin:2017vwb,Kaplan:2016drz}. However, it is only tangentially related to the radiative instability discussed in this paper, even though both originate from the same dispersion relation. In particular, the infrared instability can be ignored in the ultra-relativistic limit $\omega\gg k_\bot\gg |\sigma_\chi|$ because equation 
\ball{a15}
k_z\approx \omega -\frac{1}{2\omega}\left(k_\bot^2-\lambda\sigma_\chi \omega\right)
\gal
has only  real solutions.

%%%%
\section{Fermion wave function} \label{sec:b} % p.19 of notes

The free fermion wave function $\psi$ at high energy $\e\gg p_\bot,m$ can be obtained using the same procedure. Since it satisfies the Dirac equation we are looking for a solution in the form 
\ball{b29}
\psi = \frac{1}{\sqrt{2\e V}}u(p) \phi e^{i\e z-i \e t}\,, 
\gal
where $u(p)$ is a spinor describing a free fermion  with momentum $p$ and $\phi$ is a scalar function of coordinates.   Substituting $\psi$ into  $(\partial^2+m^2)\psi=0$ and neglecting $\partial_z\phi$ compared to $\e \phi$ one obtains
\ball{b31}
2i\e\partial_z\phi +\nabla_\bot^2\phi = m^2\phi\,
\gal
with a solution
\ball{b33}
\phi= \exp\left\{ i\b p_\bot\cdot \b x_\bot-iz\frac{\b p_\bot^2+m^2}{2\e}\right\}\,.
\gal
Thus, the fermion wave function is
\ball{b35}
\psi = \frac{1}{\sqrt{2\e V}}u(p)  e^{i\e z-i \e t}\exp\left\{ i\b p_\bot\cdot \b x_\bot-iz\frac{\b p_\bot^2+m^2}{2\e}\right\}\,.
\gal

%%%%%%%%%%
\section{Photon radiation}\label{sec:d}

%pp. 24,25, 

Modification of the photon dispersion relation in chiral matter makes possible spontaneous photon radiation $f(p)\to f(p')+\gamma(k)$. The corresponding scattering matrix element reads 
\bal
S=& -ie Q\int \bar \psi \gamma^\mu \psi A_\mu d^4x\label{d3}\\
=&-ie Q(2\pi)\delta(\omega+\e'-\e)\frac{\bar u (p')\gamma^\mu u(p)\epsilon^*_\mu}{\sqrt{8\e\e' \omega}}\int_{-\infty}^\infty dz\int d^2x_\bot\, \phi_{p'}^*(z,\b x_\bot)\varphi^*_k(z,\b x_\bot)\phi_{p}(z,\b x_\bot) \label{d4}\\
=&i(2\pi)^3\delta(\omega+\e'-\e)\delta(\b p_\bot-\b k_\bot-\b p'_\bot)\frac{\mathcal{M}}{\sqrt{8\e\e' \omega V^3}}\,,
\label{d5}
\gal
where $Q$ is the fermion electric charge.  The wave functions $\varphi_k$ and $\phi_p$ are given by \eq{b22} and \eq{b35} respectively with the subscripts indicating the corresponding momenta. The amplitude  $\mathcal{M}$ is given by 
\bal
\mathcal{M}=& -eQ\bar u(p')\gamma^\mu u(p)\epsilon^*_\mu\int_{-\infty}^\infty dz \exp\left\{ i\int_0^z dz' \left[ \frac{{p'}^2_\bot+m^2}{2\e'}-\frac{p_\bot^2+m^2}{2\e}+\frac{k_\bot^2-\sigma_\chi\omega\lambda}{2\omega}\right]\right\}
\label{d7}\\
=& \mathcal{M}_0\int_{-\infty}^\infty dz \exp\left\{ i\int_0^z  \frac{q_\bot^2+ \kappa_\lambda(z')}{2\e x(1-x)}dz'\right\} \,,
\label{d8}
\gal 
where we introduced notations $\mathcal{M}_0=-eQ\bar u(p')\gamma^\mu u(p)\epsilon^*_\mu$, $x=\omega/\e$,
\ball{d10}
\b q_\bot = x\b p'-(1-x)\b k_\bot\,,
\gal
and 
\ball{d12}
\kappa_\lambda(z) = x^2m^2-(1-x)x\lambda \sigma_\chi \e\,.
\gal
%
% pp. 30-32
%
The amplitude $\mathcal{M}_0$ can most efficiently be computed in the helicity basis using the matrix elements derived in \cite{Lepage:1980fj}. Keeping in mind that at high energies $k^+= xp^+$ (where $p^+= \e+p_z$, $k^+= \omega+k_z$), one obtains
\bal
\mathcal{M}_0&=-eQ\bar u_{\sigma'}(p')\gamma\cdot \epsilon^*_{\lambda}(k) u_{\sigma}(p) \label{d28}\\
&= -\frac{eQ}{\sqrt{2(1-x)}}\left[ xm(\sigma+\lambda)\delta_{\sigma',-\sigma}
-\frac{1}{x}(2-x+x\lambda \sigma)(q_x-i\lambda q_y)\delta_{\sigma',\sigma}
\right]\label{d29}\,,
\gal
where $\sigma=\pm 1$ and $\sigma'\pm 1$ are the fermion helicities before and after photon radiation. 

The transition probability can be computed as
\ball{d33}
dw= |S|^2\frac{V d^3p'}{(2\pi)^3}\frac{V d^3k}{(2\pi)^3}= |S|^2\frac{V d^2p_\bot dp_z'}{(2\pi)^3}\frac{V d^2q_\bot dk_z}{(2\pi)^3}
\gal
The cross section is the rate per unit flux $V^{-1}$, while the number of produced photons $N$  is the cross section per unit area. Using the usual rules for dealing with the squares of the delta-functions and integrating over the phase space yields
\ball{d35}
dN= \frac{1}{(2\pi)^3}\frac{1}{8x(1-x)\e^2}\frac{1}{2}\sum_{\lambda,\sigma,\sigma'}|\mathcal{M}|^2d^2q_\bot dx\,,
\gal
where the sum runs over the photon and fermion helicities. Eqs.~\eq{d35},\eq{d29},\eq{d8} give the spectrum of radiated photons. In the following subsections the explicit expressions for the photon spectrum are derived for a single domain and for two domains separated by a domain wall at $z=0$. 

%%%%
\subsection{One infinite domain}

Consider first an infinite chiral matter with constant chiral conductivity. Performing  the integral over $z$ in \eq{d8} yields
\ball{da1}
\mathcal{M}&= 2\pi \mathcal{M}_0\,\delta\left(\frac{q_\bot^2+\kappa_\lambda}{2\e x(1-x)}\right)\,.
\gal
The square of the delta-function in \eq{da1} is interpreted  as the delta-function multiplied by $T/2\pi$, where $T$ is  the observation time. Namely,
\ball{da2}
|\mathcal{M}|^2= 4\pi\e x(1-x)\delta(q_\bot^2+\kappa_\lambda)T|\mathcal{M}_0|^2\,. 
\gal
The  relevant intensive observable quantity then is the photon radiation rate $W$ given by
\ball{d50}
\frac{dW}{dx}= \frac{1}{16\pi\e}\frac{1}{2}\sum_{\lambda,\sigma,\sigma'}|\mathcal{M}_0|^2\,\theta(-\kappa_\lambda)\,,
\gal
where $\theta$ is the step-function. It follows from \eq{d12} that $\kappa_\lambda$ is negative if $\lambda\sigma_\chi>0$ and 
\ball{d52} 
x<x_0=\frac{1}{1+m^2/(\lambda\sigma_\chi\e)}\,.
\gal 
Assume for definitiveness that $\sigma_\chi>0$. Then only the right-polarized photons with $\lambda>0$ are radiated. Using \eq{da1}, \eq{d29} in \eq{d50} and performing the  summations and the integration yields the density of spontaneously radiated photons
\bal
\frac{dW_+}{dx}&= \frac{\alpha Q^2}{2\e x^2(1-x)}\left\{ -\left(\frac{x^2}{2}-x+1\right)\kappa_++\frac{x^4m^2}{2}\right\}\theta(x_0-x)
\nonumber\\
&= \frac{\alpha Q^2}{2\e x }\left\{ \sigma_\chi \e \left(\frac{x^2}{2}-x+1\right)-m^2x\right\}\theta(x_0-x)\,,
\label{d55}\\
\frac{dW_-}{dx}&=0\,.\label{d56}
\gal
Photon spectrum radiated in a matter with $\sigma_\chi<0$ can be obtained by  replacing $W_\pm\to W_\mp$ and $\sigma_\chi \to -\sigma_\chi$. Note that since the anomaly coefficient $ c_A\sim  \alpha$, the spectrum \eq{d55} is of the order  $\alpha^2$. 

The total energy radiated by a fermion per unit time is
\ball{d58}
\frac{\Delta \e}{T} = \int_0^1 \frac{dW_+}{dx} x\e dx= \frac{1}{3}\alpha Q^2\sigma_\chi \e\,,
\gal
where the terms of order $m^2/|\sigma_\chi|\e$ have been neglected for simplicity. Thus, energy loss increases exponentially with time. It can be neglected only for time intervals much smaller than $ \sim 1/|\sigma_\chi|\alpha$. 

%%%%
\subsection{Two semi-infinite domains separated by a domain wall at $z=0$}\label{sec:e2}

% p.26
Suppose now that the chiral matter consist of two  semi-infinite domaines separated by a thin domain wall at $z=0$.  Performing  the integral over $z$ in \eq{d8} yields  
\bal
\mathcal{M}&= \mathcal{M}_0 \left\{ \int_{-\infty}^0 dz \,e^{iz \frac{q_\bot^2+\kappa'_\lambda-i\delta}{2\e x(1-x)}}
+\int_0^{\infty} dz\, e^{iz \frac{q_\bot^2+\kappa_\lambda+i\delta}{2\e x(1-x)}}
\right\}
\label{d25}\\
&=  2\e x(1-x)\mathcal{M}_0\left\{ \frac{-i}{q_\bot^2+\kappa'_\lambda-i\delta}-\frac{-i}{q_\bot^2+\kappa_\lambda+i\delta}
\right\}\,, \label{d26}
\gal
where the values of $\kappa_\lambda$ at $z<0$ and $z>0$ are denoted by $\kappa'_\lambda$ and $\kappa_\lambda$ respectively and $\delta>0$ is inserted to regularize the integrals.  Plugging \eq{d26}, \eq{d29} into \eq{d35} and performing summation over spins yields the radiation spectrum 
% p.32,33
\ball{d40}
\frac{dN}{d^2q_\bot dx}= \frac{\alpha Q^2}{2\pi^2 x}\left\{ \left(\frac{x^2}{2}-x+1\right)q_\bot^2+\frac{x^4m^2}{2}\right\}
\sum_\lambda \left| \frac{1}{q_\bot^2+\kappa'_\lambda-i\delta}-\frac{1}{q_\bot^2+\kappa_\lambda+i\delta}\right|^2\,.
\gal
The spectrum peaks at $q_\bot^2=-\kappa_\lambda$ and/or  $q_\bot^2=-\kappa'_\lambda$ provided that $\kappa_\lambda<0$ and/or  $\kappa_\lambda'<0$ respectively. In the limit $\kappa_\lambda\to \kappa'_\lambda$  
the results of the previous subsection, provided that the square of the delta functions is treated as explained after \eq{da1}. Let us also note that when $ q_\bot^2+\kappa_\lambda=0$ in \eq{d25}, the second integral equals $T/2$, which implies that we have to identify $\delta= 4\e x(1-x)/T$ (the same result is of course obtained using the first integral).

Away from the poles, one can neglect $\delta$ in \eq{d40}. The resulting spectrum coincides with the spectrum of the transition radiation once $\kappa_\lambda$'s are replaced by $\kappa_\mathrm{tr}=m^2x^2+m_\gamma^2(1-x)$, where  $m_\gamma$ is the effective photon mass   \cite{Baier:1998ej,Schildknecht:2005sc}. Unlike the spontaneous radiation, the transition radiation is not possible in a uniform matter. Indeed,  the amplitude \eq{da1} vanishes because $\kappa_\mathrm{tr}>0$. Another key difference between the transition and spontaneous radiation is that the former has a finite classical limit $\hbar \to 0$, while the later one does not. The spontaneous radiation spectrum  \eq{dc1},\eq{dc2} is a purely quantum effect that vanishes in the classical limit $\hbar\to 0$. This is of course not surprising at all because it originates from a quantum anomaly. 

Integral over the momentum $q_\bot$ in \eq{d40} is dominated by the poles at $q_\bot^2=-\kappa_\lambda$ and $q_\bot^2=-\kappa'_\lambda$. There are two distinct cases depending on whether $\sigma_\chi$ and $\sigma_\chi'$ have the same or opposite signs. Consider first $\sigma_\chi>0$ and $\sigma_\chi'>0$. In this case the photon spectrum is approximately right-polarized. Keeping only the terms proportional to $1/\delta$ one obtains
% see transit-1.nb
\ball{dc1}
\frac{dW_{++}}{dx}= \frac{\alpha Q^2}{8 x^2(1-x)\e}\left[ \left(\frac{x^2}{2}-x+1\right)|\kappa_++\kappa_+'|+\frac{x^4m^2}{2}\right]
\theta(x_0-x)\theta(x_0'-x)\,,
\gal
where the double plus subscript indicates that the helicity is positive in both domains.
The maximum energy fraction taken by the photon $x_0$ is defined in \eq{d52}; $x_0'$ is the same as $x_0$ with $\sigma_\chi$ replaced by $\sigma_\chi'$.   Consider now  $\sigma_\chi'>0$ and $\sigma_\chi<0$. The integration gives
\ball{dc2}
\frac{dW_{+-}}{dx}=& \frac{\alpha Q^2}{8 x^2(1-x)\e}\left\{ \left[\left(\frac{x^2}{2}-x+1\right)|\kappa_+'|+\frac{x^4m^2}{4}
\right] \theta(x_0'-x)\right. \nonumber\\
&
\left.
+\left[\left(\frac{x^2}{2}-x+1\right)|\kappa_-|+\frac{x^4m^2}{4}
\right] \theta(x_0-x)\right\}\,.
\gal
Clearly,  photons radiated to the left of the domain wall ($z<0$)  are right-polarized, while those radiated to its right ($z>0$) are left-polarized.  

%%%%%%%%%%
%%%%%%%%%%
%%%%%%%%%%
\section{Pair production}\label{sec:e}

% pp. 34-39
Momentum conservation prohibits the spontaneous photon decay $\gamma(k)\to \bar f (p)+f(p')$ in vacuum. However, in chiral matter this channel is open due to the chiral anomaly. This is the cross-channel of the photon radiation computed in the previous section. The scattering matrix is now given by 
\ball{e1}
S= i(2\pi)^3\delta(\omega-\e'-\e)\delta(\b k_\bot-\b p_\bot-\b p'_\bot)\frac{\mathcal{M}}{\sqrt{8\e\e' \omega V^3}}\,,
\gal
where 
\ball{e3}
\mathcal{M}=& -eQ\bar u(p')\gamma^\mu v(p)\epsilon_\mu\int_{-\infty}^{+\infty} dz \exp\left\{ i\int_0^z  \frac{\tilde q_\bot^2+ \tilde \kappa_\lambda(z')}{2\omega x(1-x)}dz'\right\} \,.
\gal
We introduced new notations $x= \e'/\omega$, 
\ball{e5}
\tilde{\b q}_\bot = \b p'_\bot-x\b k_\bot\,,
\gal
and 
\ball{e7}
\tilde\kappa_\lambda(z) = m^2+(1-x)x\lambda \sigma_\chi \omega\,.
\gal
Notice that the sign in front of the second term is opposite to that of \eq{d12}. Using the matrix elements listed in  \cite{Lepage:1980fj} one obtains
\bal
\mathcal{M}_0&=-eQ\bar u(p')\gamma^\mu v(p)\epsilon_\mu \label{e9}\\
&= -\frac{eQ}{\sqrt{2x(1-x)}}\left[- m(\sigma+\lambda)\delta_{\sigma',\sigma}
+(2x-1-\lambda \sigma)(\tilde q_x+i\lambda \tilde q_y)\delta_{\sigma',-\sigma}
\right]\label{e10}\,.
\gal

%%%%
\subsection{One infinite domain}

In the case the entire matter is a single domain with constant $\sigma_\chi$, integration over $z$ produces the delta function similar to the one in \eq{da1}. Substituting the amplitude \eq{e3} into \eq{d35} and performing summation over spins yields the photon decay rate 
\ball{e25}
\frac{dW}{dx}= \frac{\alpha Q^2}{4x(1-x)\omega}\left\{ \left(x^2+(1-x)^2\right)(-\tilde \kappa)+m^2\right\}\theta(-\tilde \kappa)\,.
\gal
The condition $\tilde\kappa<0$ is satisfied if $\lambda \sigma_\chi<0$, $\omega>4m^2/|\sigma_\chi|$ and $x_1<x<x_2$ where
\ball{e27}
x_{1,2}=\frac{1}{2}\left( 1\mp \sqrt{1-\frac{4m^2}{|\sigma_\chi|\omega}}\right)\,.
\gal
Thus, if $\sigma_\chi>0$, then only left-polarized photons with $\lambda=-1$ can produce a pair, whereas the right-polarized photons cannot decay at all. The corresponding spectrum is 
\bal
\frac{d W_-}{dx}&= \frac{\alpha Q^2}{4\omega}\left\{ \left(x^2+(1-x)^2\right)\sigma_\chi\omega+2m^2 \right\}\theta(x_2-x)\theta(x-x_1)\,,\label{e29}\\
\frac{d W_+}{dx}&=0\,. \label{e30}
\gal
In a domain with $\sigma_\chi<0$ only right-polarized photons decay. The corresponding spectrum is obtained by replacing $W_\pm\to  W_\mp$ and $\sigma_\chi \to -\sigma_\chi$ in \eq{e29} and \eq{e30}.

%%%%
\subsection{Two semi-infinite domains separated by a domain wall at $z=0$}

The calculation in the case of chiral matter consisting of two semi-infinite domaines separated by a thin domain wall at $z=0$ is analogous to that in \sec{sec:e2}. The result is
\bal
\frac{d N}{d^2\tilde q_\bot dx}&= \frac{1}{(2\pi)^3}\frac{1}{8x(1-x)\omega^2}\frac{1}{2}\sum_{\lambda,\sigma,\sigma'}|\mathcal{M}|^2  \label{e12}\\
&= \frac{\alpha Q^2}{4\pi^2}\left\{ \left(x^2+(1-x)^2\right)\tilde q_\bot^2+m^2\right\}
\sum_\lambda \left| \frac{1}{\tilde q_\bot^2+\tilde \kappa'_\lambda-i\delta}-\frac{1}{\tilde q_\bot^2+\tilde \kappa_\lambda+i\delta}\right|^2\,, \label{e13}
\gal
where the values of $\tilde \kappa_\lambda$ at $z<0$ and $z>0$ are denoted by $\tilde \kappa'_\lambda$ and $\tilde \kappa_\lambda$ respectively. Replacing $\tilde\kappa_\lambda\to m^2-m_\gamma^2x(1-x)$ yields the transition pair production spectrum \cite{Baier:1998ej,Schildknecht:2005sc}. 
Integration over $\tilde {\b q}_\bot$ gives 
\ball{e15}
\frac{dW_{--}}{dx}=\frac{\alpha Q^2}{16x(1-x)\omega}\left\{ \left(x^2+(1-x)^2\right)|\kappa_-'+\kappa_-| +m^2\right\}\theta(x-\max(x_1,x_1'))\theta(\min(x_2,x_2')-x)\,,
\gal
if $\sigma_\chi'>0$ and $\sigma_\chi>0$ and 
\ball{e16}
\frac{dW_{-+}}{dx}=&\frac{\alpha Q^2}{16x(1-x)\omega}\left\{ 
\left[  \left(x^2+(1-x)^2\right)|\kappa_-'| +\frac{m^2}{2}\right]\theta(x-x_1')\theta(x_2'-x)
\right. \nonumber\\
&+
\left.
\left[  \left(x^2+(1-x)^2\right)|\kappa_+| +\frac{m^2}{2}\right]\theta(x-x_1)\theta(x_2-x)
\right\}\,.
\gal
if $\sigma_\chi'>0$ and $\sigma_\chi<0$. Here $x'_{1,2}= 1/2\mp \sqrt{1/4-m^2/|\sigma'_\chi|\omega}$. Eqs.~\eq{e15} and \eq{e16} clearly indicate that in a domain with positive/negative chiral conductivity most pairs are produced by left/right-polarized photons. 

%%%%%%%%%%
\section{Summary and discussion}\label{sec:s}

The main result of this paper is that a free charged fermion moving through chiral matter \emph{spontaneously} radiates electromagnetic radiation. This indicates instability of the single-fermion states. We derived the radiation spectrum in two cases: when the matter is a single $CP$-odd domain, given by \eq{d55}--\eq{d56},  and when it consists of two such domains separated by a flat thin domain wall, given by \eq{d40}. The photon polarization is determined by the sign of the chiral conductivity: if it is positive/negative, the radiation is right/left-polarized. The cross-channel process of spontaneous photon radiation is spontaneous pair production by a real photon. This indicates instability of the single-photon states. We computed the fermion spectrum and found that in a domain with positive/negative chiral conductivity only left/right-polarized photons decay, see \eq{e29},\eq{e30}.

The rate of energy loss by single-particle states is found to be proportional to $\alpha \sigma_\chi$. Since the temporal evolution of chiral conductivity has much higher rate of $\sigma_\chi$, it seems plausible that it can play an important role in the long-time dynamics of the radiative instability and perhaps even tame it. This is a problem that deserves further investigation. 

In electromagnetic plasma,  the existence of the $CP$-odd domains would trigger the radiative instability causing radiation of photons of a certain polarization, and decay of photons of opposite polarization. These processes tend to polarize the plasma within a domain. Since the $2\to 2$ scattering as well as transitions due to spatial inhomogeneities \cite{Tuchin:2016qww} are also of order $\alpha^2$ the question of whether there is an equilibrium polarization of electromagnetic field requires further investigation.

Despite the radiative instability, the Maxwell-Chern-Simons effective theory \eq{a3} is a useful tool to study macroscopic effects of the chiral anomaly if $\sigma_\chi$ is a sufficiently small compared to the typical energy scales. This is the case in the  quark-gluon plasma where $\sigma_\chi$ is about two order of magnitudes lower than the plasma temperature \cite{Kharzeev:2001ev,Fukushima:2012fg}. Thus, convoluting \eq{d55} with the Fermi-Dirac distribution over the phase space gives photon spectrum spontaneously radiated by quarks in quark-gluon plasma. Since the spectrum is proportional to $\alpha^2$ it gives only a minor contribution to the total photon spectrum radiated by the plasma. However, the radiative instability due to the chiral anomaly of QCD is strongly enhanced by $\alpha_s/\alpha$ and may have a  significant impact on the quark-gluon plasma phenomenology.

%%%%%%%%%%%%%%%%%%%%%%%%%%%%%%%%
\acknowledgments
This work was supported in part by the U.S. Department of Energy under Grant No.\ DE-FG02-87ER40371.

%%%%%%%%%%%%%%%%%%%%%%%%%%%%%%%%%%%%%

\end{document}